\documentclass[fleqn,usenatbib]{mnras}

\usepackage{newtxtext,newtxmath}

\usepackage[T1]{fontenc}

\DeclareRobustCommand{\VAN}[3]{#2}
\let\VANthebibliography\thebibliography
\def\thebibliography{\DeclareRobustCommand{\VAN}[3]{##3}\VANthebibliography}

\usepackage{graphicx}	% Including figure files
\usepackage{amsmath}	% Advanced maths commands
\usepackage{graphicx}	% Including figure files
\usepackage{subcaption} % Subfigures
\graphicspath{ {./images/} }

\usepackage{amsmath}	% Advanced maths commands
\usepackage{amssymb}    % Extra maths symbols
\usepackage{bm}         % Bold

\title[Flux tube clustering in NS]{Flux tube clustering from magnetic coupling of adjacent type-I and -II superconductors in a neutron star: persistent gravitational radiation}

\author[K. H. Thong and A. Melatos]{
K. H. Thong$^{1,2,3}$
\thanks{E-mail: kokhongt@student.unimelb.edu.au}
and
A. Melatos$^{1,2}$
\thanks{E-mail: amelatos@unimelb.edu.au}
\\
$^{1}$School of Physics, University of Melbourne, Parkville, Victoria, 3010 Australia\\
$^{2}$OzGrav, Australian Research Council Centre of Excellence for Gravitational Wave Discovery, University of Melbourne, Parkville, Victoria, 3010 Australia\\
$^{3}$Department of Applied Mathematics and Theoretical Physics, University of Cambridge, Centre for Mathematical Sciences, Wilberforce Rd, Cambridge \\CB3 0WA}

\date{Accepted XXX. Received YYY; in original form ZZZ}

\pubyear{2015}

\begin{document}
\label{firstpage}
\pagerange{\pageref{firstpage}--\pageref{lastpage}}
\maketitle

\begin{abstract}
Adjacent type-I and -II proton superconductors in a rotation-powered pulsar are predicted to exist in a metastable state containing macroscopic and quantized flux tubes, respectively. Previous studies show that the type-I and -II regions are coupled magnetically, when macroscopic flux tubes divide dendritically into quantized flux tubes near the type-I-II interface, through a process known as flux branching. 
%Flux branching magnetically couples the quantized flux tubes with macroscopic flux, forming a macroscopic structure known as a ``flux tree''. 
The studies assume that the normal-superconducting boundary is sharp, and the quantized flux tubes do not repel mutually. 
%which is adequate to describe branching in the type-I superconductor but does not predict the separation distance between quantized flux tubes in the same flux tree in the type-II superconductor. 
Here the sharp-interface approximation is refined by accounting for magnetic repulsion.
It is found that flux tubes in the same flux tree cluster with a minimum-energy separation two to seven times less than that of isolated flux tubes.
Neutron vortices pin and cluster about flux trees.
We find that the maximum characteristic wave strain $h_0$ of the current quadrupole gravitational radiation emitted by a rectilinear array of clustered vortices exceeds by $(1+N_{\rm v,t})^{1/2}$ the strain $h_0 \sim 10^{-32}(f/30 {\rm Hz})^{5/2} (D/1 {\rm kpc})^{-1}$ emitted by uniformly distributed vortices, where $N_{\rm v,t}$ is the mean number of pinned vortices per flux tree, $f$ is the star's spin frequency, and $D$ is the star's distance from Earth.
The factor $(1 + N_{\rm v,t})^{1/2}$ brings $h_0$ close to the sensitivity limit of the current generation of interferometric gravitational wave detectors under certain circumstances, specifically when flux branching forms relatively few (and hence relatively large) flux trees.
% This is a simple template for authors to write new MNRAS papers.
% The abstract should briefly describe the aims, methods, and main results of the paper.
% It should be a single paragraph not more than 250 words (200 words for Letters).
% No references should appear in the abstract.
\end{abstract}


%%%%%%%%%%%%%%%%%%%%%%%%%%%%%%%%%%%%%%%%%%%%%%%%%%

%%%%%%%%%%%%%%%%% BODY OF PAPER %%%%%%%%%%%%%%%%%%

\section{Introduction}

Standard, rotation-powered pulsars, with magnetic field $B \sim 10^{12} {\rm \ G} \ll H_{\rm c} \sim H_{\rm c1} \sim 10^{15} {\rm \ G}$, are believed to host type-I and -II proton superconductors in their inner and outer cores, respectively, where $H_{\rm c}$ is the critical magnetic field of the type-I superconductor and $H_{\rm{c1}}$ is the first critical magnetic field of the type-II superconductor \citep{sedrakian_type_1997,jones_type_2006,glampedakis_magnetohydrodynamics_2011,haskell_investigating_2013,haber_critical_2017}. 
Due to flux freezing, the type-I regions are threaded by macroscopic flux tubes\footnote{For simplicity, we assume in this paper that the macroscopic flux tubes are cylindrical, although realistically they depend on the magnetic evolutionary history of the star, which is hard to predict theoretically \citep{tout_magnetic_2004,ferrario_magnetic_2015}.}, while the type-II regions are threaded by quantized microscopic flux tubes \citep{baym_superfluidity_1969}. 
A recent study \citep{thong_magnetic_2024} demonstrates that if the magnetic field lines in the type-I and -II superconductors are connected, then macroscopic flux tubes in the type-I region divide dendritically into quantized flux tubes in the type-II region. This phenomenon, formally known as flux branching \citep{landau_theory_1943}, arises as the type-I and -II superconductors compete to minimize and maximize $S$, the normal-superconducting surface area, respectively. Flux branching couples quantized flux tubes in the outer core with macroscopic flux tubes in the inner core, forming a macroscopic object known as a flux tree. 

\citet{thong_magnetic_2024} assumed that the normal-superconducting interface is sharp and that flux-containing normal regions do not mutually interact, as in many previous analyses \citep{landau_theory_1943,andrew_intermediate_1948,hubert_theory_1967,choksi_energy_2004}. Consequently, the mutual repulsion of the quantized flux tubes in the type-II region is neglected. 
The sharp-interface approximation is acceptable for investigating the overall energetics of flux branching \citep{landau_theory_1943,andrew_intermediate_1948}. However, it does not lead to an accurate estimate for the minimum-energy separation between the quantized flux tubes in the outer core. 
Several interesting and related questions arise. (i) Naively, without flux branching and neglecting the proton's strong coupling to the neutron superfluid, the flux-tube separation, $ d_{\Phi, {\rm nb}}$, in the outer core is assumed to be maximized due to mutual repulsion, with $ d_{\Phi, {\rm nb}} \approx 5 \times 10^{-10} \, {\rm cm} $, where the subscript nb denotes no flux branching \citep{baym_superfluidity_1969,glampedakis_magnetohydrodynamics_2011}. Hence, with flux branching, does the flux-tube separation in the outer core reduce compared to $d_{\Phi, {\rm nb}}$, such that flux tubes are distributed inhomogeneously in clusters in the outer core?
(ii) Does flux branching create an effective attractive length-scale for quantized flux tubes in the same flux tree, in addition to the attractive length-scale found in the outer core due to strong coupling between the proton superconductor and the neutron superfluid \citep{haber_critical_2017,wood_superconducting_2022}?
% Several interesting and related questions arise. 
% Is the flux-tube separation in the outer core less with flux branching than without?
% If so, does flux branching create an additional attractive length-scale for quantized flux tubes in the same flux tree, such that flux tubes are distributed inhomogeneously in clusters in the outer core, like those found in type-1.5 superconductivity \citep{moshchalkov_type-15_2009,silaev_microscopic_2011,haber_critical_2017,wood_superconducting_2022}? 
(iii) Flux branching in a type-I superconductor immersed in a weak applied magnetic field is predicted to cause inhomogeneous clustering of flux \citep{choksi_energy_2004}. Does this phenomenon extend to flux branching in adjacent type-I and -II superconductors? 

\begin{table}
	\centering
	\caption{Approximate length-scales of the type-II superconductor at density $\rho \approx 2.8 \times 10^{14}$ g cm$^{-3}$ in a standard, rotation-powered pulsar with magnetic field $B \sim 10^{12} \ {\rm G}$ , proton gap $\Delta_{\rm p} \sim 1 \ {\rm MeV}$, proton fraction $x_{\rm p} \approx 0.05$, ratio of effective $(m_{\rm p}^*)$ and bare $(m_{\rm p})$ neutron mass, $m_{\rm p}^*/m_{\rm p} \approx 1/2$, and superfluid transition temperature $T_{\rm cp} \sim 10^9 \ {\rm K}$  . $\lambda, \, \xi,$ and $d_{\Phi,{\rm nb}}$ are the London penetration depth, proton coherence length, and quantized flux-tube separation without flux branching, respectively \citep{baym_superfluidity_1969,mendell_superfluid_1991,glampedakis_magnetohydrodynamics_2011,link_instability_2012}.}
	\label{table:1}
	\begin{tabular}{lccr} % four columns, alignment for each
		\hline
		Scale & Value  \\
		\hline
		$\lambda$& 
            $6 \times 10^{-12} \left(\frac{\rho}{2.8 \times 10^{14}\, {\rm g \, cm^{-3}}} \right)^{-1/2}$ cm\\
		$\xi$&  $2 \times 10^{-12}\left(\frac{\rho}{2.8 \times 10^{14}\, {\rm g \, cm^{-3}}} \right)^{1/3}$ cm \\
		$d_{\Phi,{\rm nb}}$&  $5 \times 10^{-10} \left(\frac{B}{10^{12}\, {\rm G}} \right)^{-1/2}$ cm \vspace{1 mm}\\ 
		\hline
	\end{tabular}
\end{table}

In this paper, we investigate if quantized flux tubes cluster as a result of flux branching between adjacent type-I and -II proton superconductors. 
We extend the minimum-energy calculations in \citet{thong_magnetic_2024} by including magnetic repulsion between quantized flux tubes in the type-II superconductor. 
For standard, rotation-powered neutron stars (i.e.\ not magnetars), we take the quantized flux tubes to be widely separated with separation $d_{\Phi, {\rm nb}} \gg \lambda$, where $\lambda$ is the density-dependent London penetration depth which takes the value $\lambda \sim 6\times 10^{-12} \, {\rm cm}$ in the type-II superconductor \citep{link_instability_2012}; see the summary of characteristic length scales in Table \ref{table:1}.
In this regime, the magnetic repulsion of quantized flux tubes in a type-II superconductor can be modelled by London's theory, where the flux tube cores can be neglected \citep{tinkham_introduction_2004,glampedakis_magnetohydrodynamics_2011}.
The paper is structured as follows. In Section \ref{sec:2}, we briefly review the energetics of a flux tree within the sharp-interface approximation and demonstrate that the energy decreases with the width of the tree. 
We then refine the sharp-interface approximation by accounting for the mutual repulsion between the quantized flux tubes in the type-II superconductor. 
In Section \ref{sec:3}, we minimize the refined energy of a flux tree with respect to its width and find the preferred flux-tube separation. 
In Section \ref{sec:4}, we apply the results in an astrophysical setting to calculate the maximum current quadrupole gravitational radiation emitted by a rectilinear array of neutron superfluid vortices pinned to clustered flux tubes in a neutron star.

\section{Flux tree energetics in a type-I superconductor}
\label{sec:2}
Type-I and -II proton superconductors are likely to be adjacent in a neutron star, and coexist with and strongly couple to a neutron superfluid \citep{sedrakian_type_1997,glampedakis_magnetohydrodynamics_2011}. The coupling to the superfluid leads to interesting modifications of the superconducting phase diagram, a different set of critical fields, and an inhomogeneous, non-metastable, non-Meissner equilibrium configuration \citep{haber_critical_2017,wood_superconducting_2022}. We neglect these complexities in this paper. That is, we assume that the type-I and -II superconductors exist in Meissner states in equilibrium, where the flux is completely expelled.
However, due to the high electrical conductivity of the core, flux expulsion takes longer than the age of the star \citep{baym_electrical_1969}, so that equilibrium is not reached; instead, the superconductors are in metastable states threaded by flux. 
The surface energy per unit area, $\gamma$, of the normal-superconducting boundary satisfies $\gamma>0$ and $\gamma<0$ in type-I and -II superconductors respectively.
Consequently, flux threads a type-II superconductor with quantized flux tubes of width $\sim \xi$ and magnetic flux quantum $\Phi_0 = \pi \hbar/2e = 2.07 \times 10^{-7} \text{ G cm}^2$, where $\xi$ is the proton coherence length, while flux threads a type-I superconductor with macroscopic flux tubes of width $\gg \xi$ and flux $\gg \Phi_0$.

Near the type-I-II interface, $\gamma$ is small, and macroscopic flux tubes divide dendritically into quantized tubes, forming a macroscopic structure called a flux tree \citep{thong_magnetic_2024} .
In Section \ref{sec:2.1}, we briefly review the flux branching formalism used by \citet{thong_magnetic_2024} to obtain the free energy of a flux tree in the type-I superconductor as a function of its width.
In Section \ref{sec:2.2}, we discuss the mutual repulsion between the quantized flux tubes in the type-II superconductor, and how this modifies the total energy of a flux tree as a function of its width.

\subsection{Sharp-interface approximation}
\label{sec:2.1}

\begin{figure}
    \centering
\begin{subfigure}[b]{0.5\textwidth}
    \centering
    \includegraphics[width=0.8\textwidth]{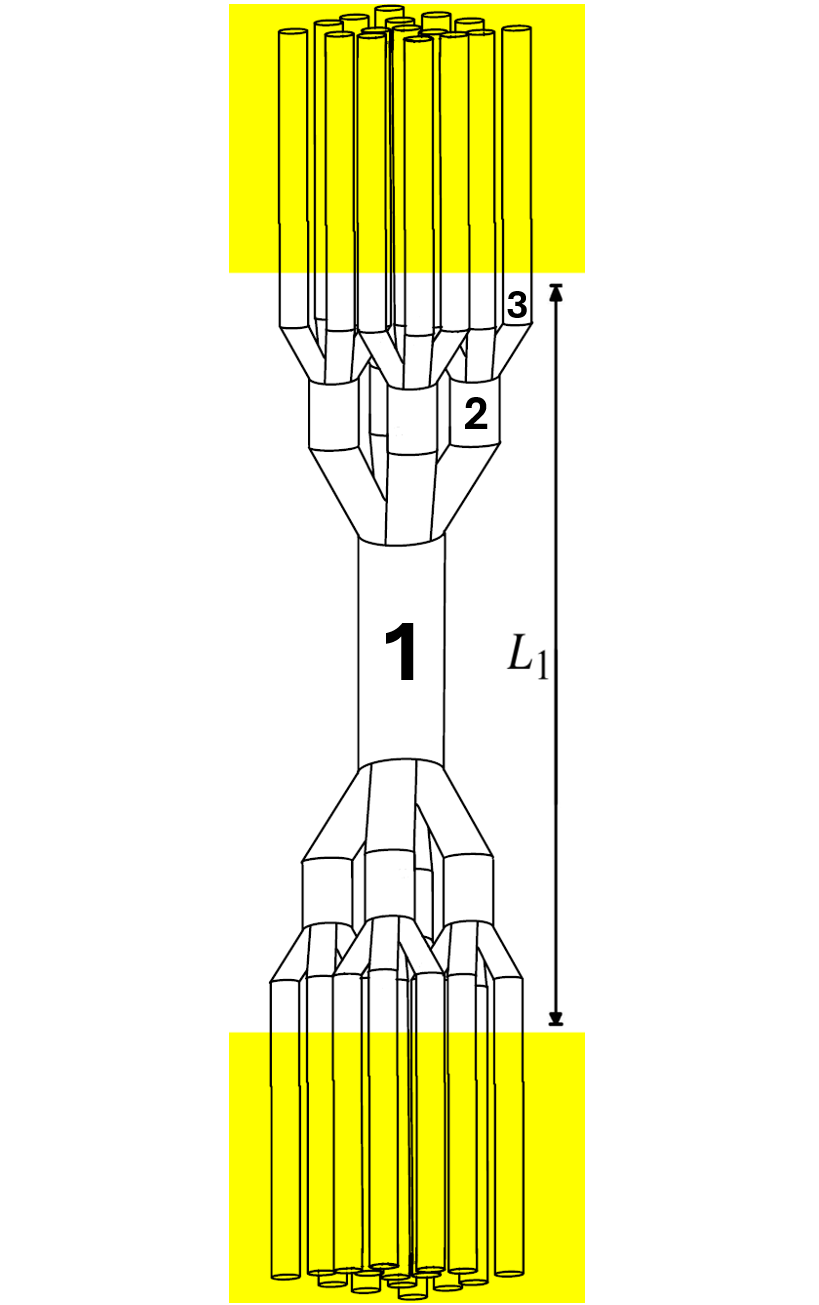}
    	\caption{}
        \label{fig:1a}
    \end{subfigure}
    \hfill
    \vspace{3mm}
    \begin{subfigure}[b]{0.5\textwidth}
    	\includegraphics[width=\textwidth]{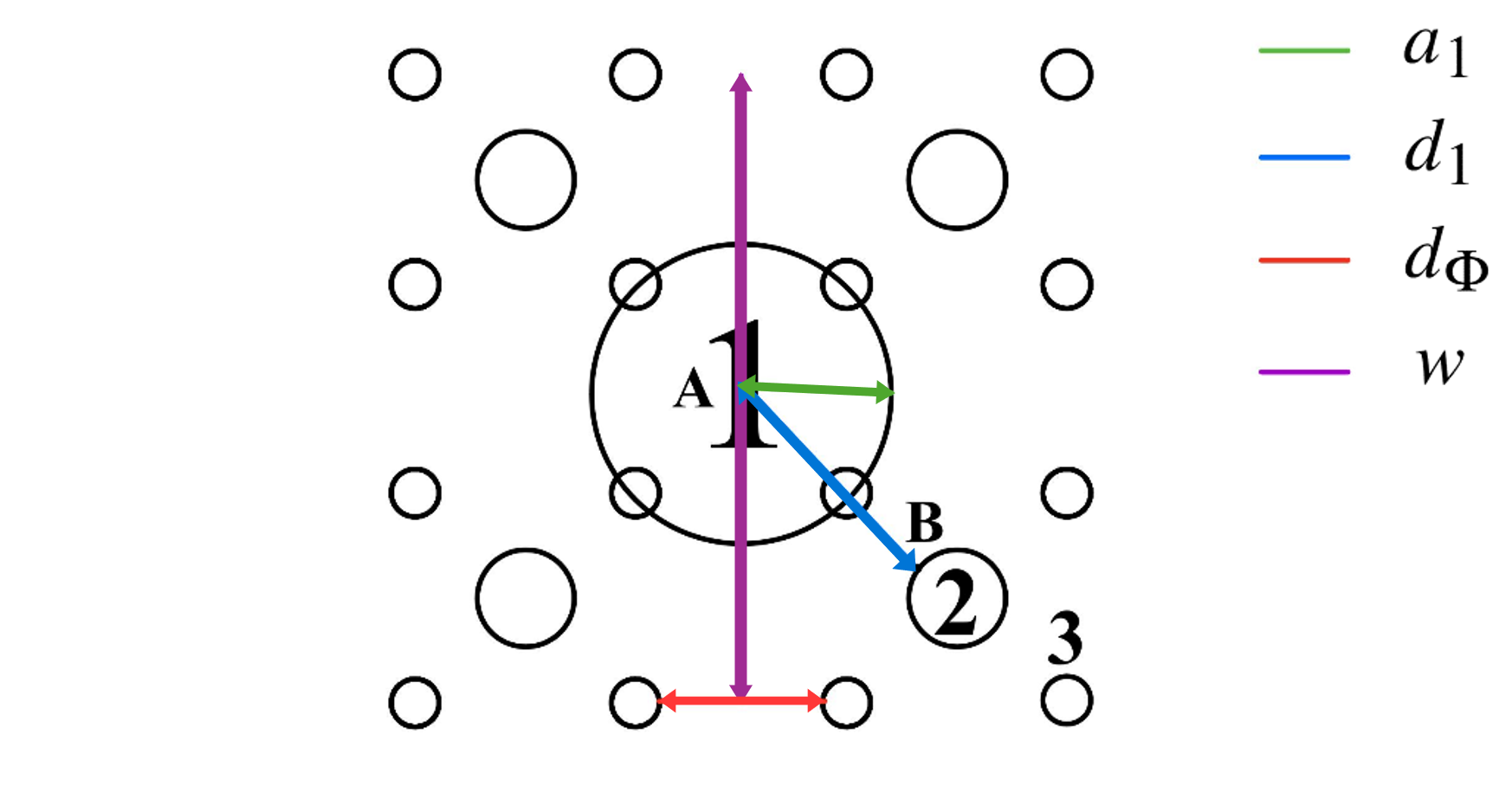}
    	\caption{}
    	\label{fig:1b}
    \end{subfigure}
    \hfill
    %\vspace{3mm}
    \caption{Schematic diagrams of a $N=3$ flux tree in adjacent type-I (no shading) and type-II (yellow shading) superconductors, adapted from \citet{andrew_intermediate_1948}. (a) Lateral view. (b) Top view. The numbers $1$--$3$ label the three different flux tubes. Quantized flux tubes (labelled by $3$) carrying a single flux quantum begin from the type-I region, cross the type-I-II interface and extend into the type-II region. $L_1$ is the length of the flux tree, $a_1$ (green line) is the radius of the largest macroscopic flux tube, $d_1$ (blue line) is the distance between the points labelled $A$ and $B$, $w$ (orange line) is the width of the flux tree and $d_\Phi$ (red line) is the separation between the smallest flux tubes.}
    \label{fig:1}
\end{figure}

\citet{thong_magnetic_2024} assumed that the normal-superconducting interface is sharp, and flux branching occurs near the type-I-II interface in the type-I superconductor (i.e., not in the type-II superconductor), following previous literature \citep{landau_theory_1943,andrew_intermediate_1948,huebener_magnetic_2001,choksi_energy_2004}.
That is, the superconducting regions have magnetic field $B=0$, while the normal regions have uniform $B = H_{\rm c}$. 
This standard approximation is valid, when $\xi$ and $ \lambda$ are much less than the size of the normal and superconducting regions \citep{chapman_asymptotic_1995,choksi_energy_2004}.

The flux trees considered by \citet{thong_magnetic_2024} have flux tubes of $N$ different radii and $N-1$ branching levels. The radii of the flux tubes are labelled by $a_n$, with $1 \leq n \leq N$. At the $n$-th branching level, there are $4^{n-1}$ branching cells, each containing a flux tube of radius $a_n$, which splits into four smaller flux tubes of radius $a_{n+1}$ with $a_n^2 = 4 a^2_{n+1}$ by flux conservation.
Successive branching cells are concatenated together, such that each flux tube with radius $a_{n+1}$ at branching level $n$ connects to flux tubes of radius $a_{n+1}$ (not $a_n$) at branching level $n+1$, in such a way as to form a square flux tree when viewed in cross-section, as in Figure \ref{fig:1b}.

To illustrate the geometry of a flux tree with $N-1$ branching levels, Figure \ref{fig:1} presents schematic diagrams of a $N=3$ flux tree, with distances $a_1$, $L_1$ (green), $d_1$ (blue), $d_\Phi$ (red) and $w$ (orange) labelled graphically.
$a_1$ is the radius of the largest macroscopic flux tube (trunk), $L_1$ is the length of the flux tree in the direction of the magnetic axis, denoted by $\vec{B}$, $w$ is the width of the flux tree, and $d_1$ is the distance between the points labelled $A$ and $B$ in Figure \ref{fig:1b}.
Given a branching cell at branching level $n$ with flux tubes characterized by $a_n$ and $a_{n+1}$, it is easy to infer from Figure \ref{fig:1b} that $d_n$ is the minimum distance from the centre of $a_n$ to the boundary of $a_{n+1}$, which is a key branching parameter related to the branch opening angle. 
Following \citet{landau_theory_1943} and \citet{andrew_intermediate_1948}, we let the opening angle halve at every successive branching level. i.e., $d_n = d_1/2^{n-1}$, ensuring that the flux tubes do not overlap.
One can infer diagrammatically [see Appendix B of \citet{thong_magnetic_2024}] that $d_1$ relates to the width, $w$, of the flux tree through
\begin{equation}
\label{eq:w}
    w = \sqrt{2} (2d_1 + a_1),
\end{equation}
and relates to the characteristic separation, $d_\Phi$, between the quantized flux tubes (smallest branches) at the type-I-II interface through
\begin{align}
\label{eq:d_Phi}
    d_\Phi =& 2^{3/2-N}\left(2d_1 + a_1 \right) \\ =& 2^{1-N} w.
\end{align}
Moreover, $d_1$ satisfies the constraint
\begin{equation}
\label{eq:d_1_constraint}
    d_1 \geq \frac{(\sqrt{2}-1) a_1}{2},
\end{equation}
to prevent the flux tubes from overlapping.
We note that the branching configuration leading to a square flux tree in \citet{thong_magnetic_2024} is only one possibility out of many, chosen simply out of convenience, and further studies are required to investigate what configuration is favored energetically for any specific evolutionary history of the star.

Assuming that the normal-superconducting interface is sharp, the free energy of a flux tree with $N$ branching levels in a type-I superconductor comprises several terms. 
The dominant term that depends on $d_\Phi$ (and by extension, $w$ and $d_1$), $F_d$, is given by \citep{thong_magnetic_2024} \footnote{Equation (\ref{eq:F_d}) equals double the corresponding term in \citet{thong_magnetic_2024} , because the latter reference analyzes the top half of a flux tree, while we consider the entire tree.} 
\begin{align}
\label{eq:F_d}
    F_d \approx \ &\frac{4H_{\rm c}^2a_1^2 d_1^2}{L_1}.
\end{align}
From (\ref{eq:F_d}), we see that $F_d$ is minimized by $d_1 = (\sqrt{2}-1)a_1/2$ \citep{thong_magnetic_2024} .
That is, a flux tree is as thin as possible, and both $w$ and $d_\Phi$ are minimized. 
However, the calculation does not account for the mutual repulsion between the quantized flux tubes (terminating smallest branches) which extend into the type-II superconductor, as explained in Section \ref{sec:2.2}.

\subsection{Interactions between quantized flux tubes in a type-II superconductor }
\label{sec:2.2}
Quantized flux tubes exhibit mutual magnetic repulsion at a distance $\sim \lambda$ and attraction at a distance $\sim \xi$. 
Attraction occurs because the superconducting condensation energy decreases, when the cores of flux tubes overlap. 
Thus, quantized flux tubes are mutually repulsive in type-II superconductors with $\kappa > 1/\sqrt{2}$ and attractive in type-I superconductors with $\kappa < 1/\sqrt{2}$, where $\kappa = \lambda/\xi$ is the Ginzburg-Landau parameter. 
Near the type-I-II interface, one has $\kappa \approx 1/\sqrt{2}$ and the quantized flux tubes interact negligibly \citep{kramer_thermodynamic_1971, tinkham_introduction_2004}. 

In this paper, we are interested in modelling the regions in the outer core where magnetic repulsion dominates the interactions between quantized flux tubes, e.g.\ the regions far away from the type-I-II interface in the type-II superconductor.
One has $B \ll H_{\rm c1}$ for a neutron star, so quantized flux tubes in the type-II superconductor are typically widely separated, with $5 \times 10^{-10} \, {\rm cm} \approx d_{\Phi, {\rm nb}} \gg \lambda \approx 6\times 10^{-12} \, {\rm cm} $ \citep{baym_superfluidity_1969, link_instability_2012}, where the subscript ${\rm nb}$ denotes no flux branching.
To calculate $d_\Phi$ from (\ref{eq:d_Phi}), we add the free energy, $F_{\rm int}$, of magnetic repulsion between quantized flux tubes in the type-II superconductor to (\ref{eq:F_d}).
For $d_\Phi \gg \lambda$, the magnetic repulsion can be approximated by solutions of the London equations \citep{tinkham_introduction_2004}. The interaction free energy, $F_{\rm int}$, of a lattice of rectilinear and widely separated quantized flux tubes is given by
\begin{equation}
\label{eq:F_int}
    F_{\rm int} = \frac{\Phi_0^2L_2}{8 \pi^2 \lambda^2} \sum_{i > j} K_0\left(\frac{r_{ij}}{\lambda}\right),
\end{equation}
where $i, j$ are lattice site indices, $L_2$ is the length of the quantized flux tubes, $r_{ij}$ is the distance between the $i$-th and $j$-th sites, and $K_0$ is the zeroth-order modified Bessel function of the second kind \citep{tinkham_introduction_2004,glampedakis_magnetohydrodynamics_2011}. 
$K_0(x)$ decreases exponentially for $x > 1$, so we approximate (\ref{eq:F_int}) by considering interactions between nearest neighbours only.
Specifically, $F_{\rm int}$ for a square lattice can be approximated by %\footnote{Square \citep{kramer_thermodynamic_1971} and triangular \citep{tinkham_introduction_2004} lattices are found to minimize the free energy in different circumstances with similar $F_{\rm int}$.}
\begin{equation}
\label{eq:F_int2}
    F_{\rm int} \approx \frac{4^{N-1}L_2\Phi_0^2}{4 \pi^2 \lambda^2} \sqrt{\frac{\pi \lambda}{2 d_\Phi}} e^{-d_\Phi/\lambda},
\end{equation}
where $4^{N-1}$ is the number of quantized flux tubes in a flux tree with $N$ branching levels \citep{tinkham_introduction_2004}.
%\footnote{(\ref{eq:F_int}) is of order unity to the variational solution from \citet{clem_simple_1975} of the full Ginzburg-Landau equations for $\kappa \approx 9/4$ \citep{glampedakis_magnetohydrodynamics_2011}.}
The total width-dependent energy of a flux tree can be written as
\begin{equation}
\label{eq:F_t1}
    F_{\rm t} = F_d + F_{\rm int}.
\end{equation}

\section{Energy minimization leading to flux tube clustering}
\label{sec:3}
In this section, we minimize $F_{\rm t}$ with respect to $d_\Phi$ from (\ref{eq:F_t1}) to find the minimum-energy separation of the quantized flux tubes.
Minimizing the function in (\ref{eq:F_t1}) analytically is impossible due to its transcendental nature, so we calculate the minimum $F_{\rm t}$ numerically in Section \ref{sec:3.1}.
We find that quantized flux tubes cluster.
In Section \ref{sec:3.2}, we discuss briefly for context a related result obtained by \citet{haber_critical_2017} and \citet{wood_superconducting_2022}, whereby flux tube clustering occurs in the outer core of a neutron star, albeit due to a different mechanism.

\subsection{Minimum-energy separation}
\label{sec:3.1}

\begin{figure}
    \centering
    \includegraphics[width=0.5\textwidth]{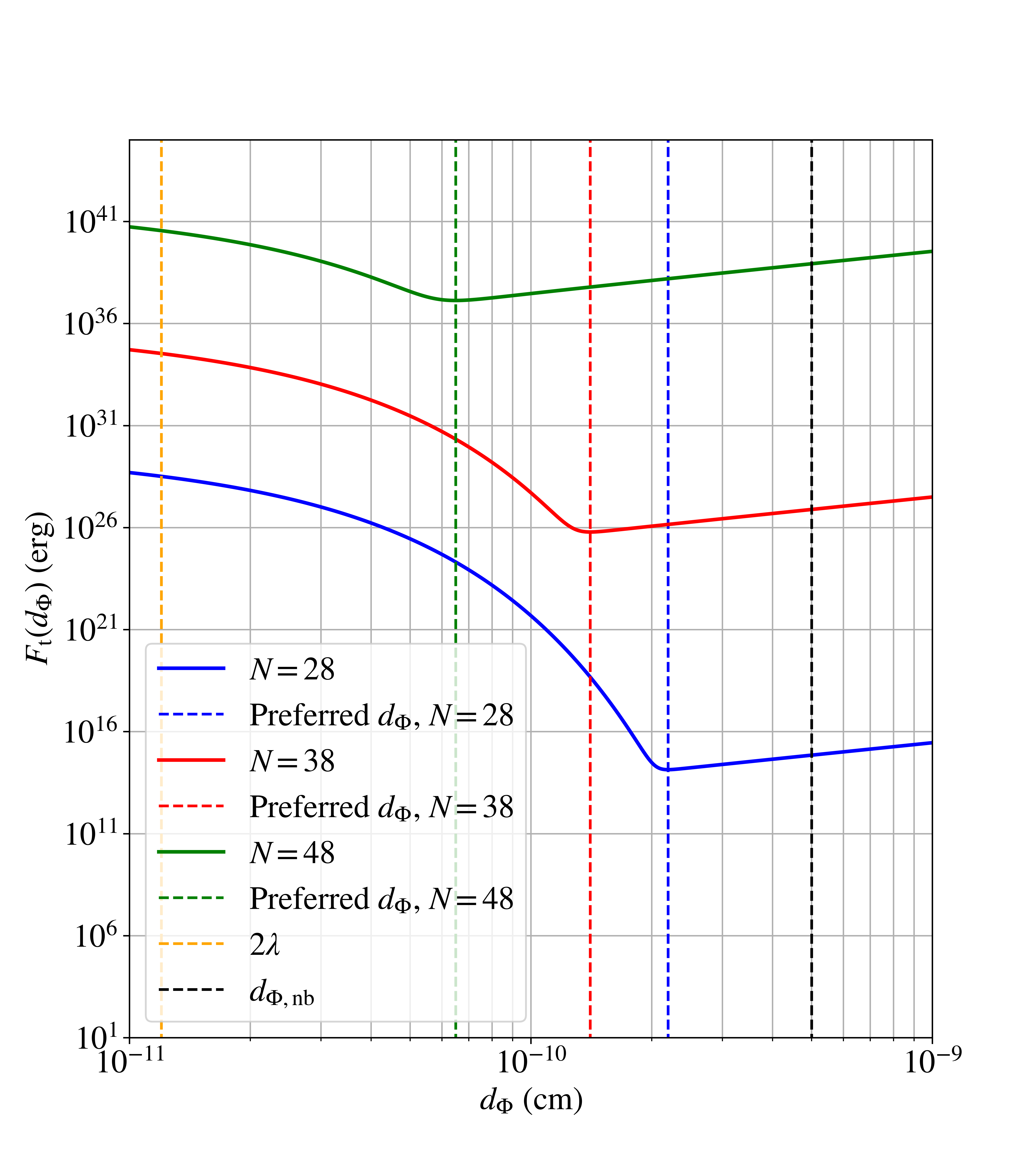}
    \caption{Width-dependent free energy, $F_{\rm t}$, of a $N$-level flux tree as a function of flux tube separation, $d_\Phi$. Blue, red and green solid curves correspond to $N = 28,\, 38$ and $48$, respectively. Dashed, color-coded, vertical lines mark the minimum of each curve, i.e.\ the minimum-energy $d_{\Phi}$. The orange and black dashed lines represent $2\lambda$ and $d_{\Phi ,{\rm nb}}$, respectively, where $2\lambda$ is the minimum separation at which magnetic repulsion becomes significant., and $d_{\Phi{,\rm nb}}$ is the separation without flux branching. Parameters: $L_1 = L_2 = 10^6 \, {\rm cm},\, H_{\rm c} = 10^{15} \, {\rm G},\, \lambda = 6 \times 10^{-12} \, {\rm cm}$ \citep{baym_superfluidity_1969}. }
    \label{fig:2}
\end{figure}
In Figure \ref{fig:2}, we plot $F_{\rm t}$ as a function of $d_\Phi$, for $N = (28, 38, 48)$ and $(L_1, L_2) / (10^{6} \, {\rm cm}) = (1, 1)$, which corresponds to the radial thicknesses of the type-I and -II regions being equal, where $L_1 + L_2 = 2 \times 10^6 \ {\rm cm}$ is the diameter of the core.
The branching level, $N-1$, governs the radius of the largest macroscopic flux
tube (trunk) of a flux tree, with $a_1 = 2^{N-1} \lambda$, as well as the number, $\mathcal{N} = 4^{N-1}$, of quantized flux tubes (smallest branches) per tree. Following \citet{thong_magnetic_2024}, we study a representative range of special cases, where $N = 28$ and $N=38$ bracket roughly the order-of-magnitude estimates in the literature for a uniform interior magnetic field, viz.\ $10^{-3} \lesssim a_1/{\rm cm} \lesssim 10^{-1}$ \citep{buckley_neutron_2004,sedrakian_type-i_2005}, and $N = 48$ corresponds to $4^{N-1} = 2 \times 10^{28}$ quantized flux tubes per flux tree, which is close to the maximum allowed. There are typically $N_{\rm f} \sim 10^{30} (B/10^{12} \, {\rm G})$ quantized flux tubes in total in the type-II superconductor, implying $N \lesssim 51$ in a standard, rotation-powered neutron star (i.e.\ not a magnetar). The values $(a_1/{\rm cm},\mathcal{N})$ are approximately $(1\times 10^{-3} ,2 \times 10^{16}), (1, 2 \times 10^{22})$ and $ (1 \times 10^3, 2 \times 10^{28})$ for $N= 28, 38$ and $48$, respectively.
We expect $a_1$ to follow some distribution set by the initial conditions (e.g.\ dynamo action) early in the star's life, before the superconductors condense, the study of which lies outside the scope of this paper \citep{tout_magnetic_2004,ferrario_magnetic_2015}.

% Following \citet{thong_magnetic_2024} , we study a representative range of special cases, where $N = 28$ and $N=38$ bracket roughly the order-of-magnitude estimates in the literature for a uniform interior magnetic field, viz.\ $10^{-3} \lesssim a_1/(1 \, {\rm cm}) \lesssim 10^{-1}$ \citep{buckley_neutron_2004,sedrakian_type-i_2005}, and $N = 48$ corresponds to $4^{N-1} = 2 \times 10^{28}$ quantized flux tubes per flux tree, which is close to the maximum allowed.
% In a neutron star, there are typically $N_{\rm f} \sim 10^{30} (B/10^{12} \, {\rm G})$ quantized flux tubes in total in the type-II superconductor, implying $N \lesssim 51$. 
% We expect $a_1$ to follow some distribution set by the initial conditions (e.g.\ dynamo action) early in the star's life, before the superconductors condense, the study of which lies outside the scope of this paper \citep{tout_magnetic_2004,ferrario_magnetic_2015}.

% In Figure \ref{fig:2}, $F_{\rm t}$ is plotted as a function of $d_\Phi$, for $N = (28, 38, 48)$ and $(L_1, L_2) / (10^{6} \, {\rm cm}) = (1, 1)$, representing equal radial thickness of the type-I and -II region.
The $d_\Phi$ values that minimize $F_{\rm t}$ for each $N$ are labelled with vertical dashed lines in Figure \ref{fig:2} and compared to $2\lambda$ and the typical flux tube separation without flux branching given by $d_{\Phi, {\rm nb}} \sim 5 \times 10^{-10} \, {\rm cm}$ \citep{baym_superfluidity_1969,glampedakis_magnetohydrodynamics_2011}.
The minimum-energy $d_\Phi$ values are calculated to be $(2.2, 1.4, 0.65) \times 10^{-10} \, \rm{cm}$ for $N=(28, 38, 48)$ respectively.
Hence (\ref{eq:w}) and (\ref{eq:d_Phi}) imply that the minimum-energy $w$ values are given by $2.9 \times 10^{-2} \, \rm{cm}, 1.9 \times 10^{2} \, \rm{cm}$ and $9.1 \times 10^{3} \, \rm{cm}$ for $N=28, 38$ and $48$ respectively. 
For $N = 38$, we also calculate the minimum-energy $d_\Phi$ for $(L_1, L_2)/ (10^{6} \, {\rm cm}) = (0.01, 1.99), (1.99, 0.01)$ to be $(1.2, 1.2) \times 10^{-10} \, \rm{cm}$, respectively.
That is, the minimum energy $d_\Phi$ and $w$ do not vary significantly with $L_1$ and $L_2$, assuming astrophysically plausible bounds on $L_1/L_2$.

\begin{figure}
    \centering
    \begin{subfigure}[b]{0.5\textwidth}
    	\includegraphics[width=\textwidth]{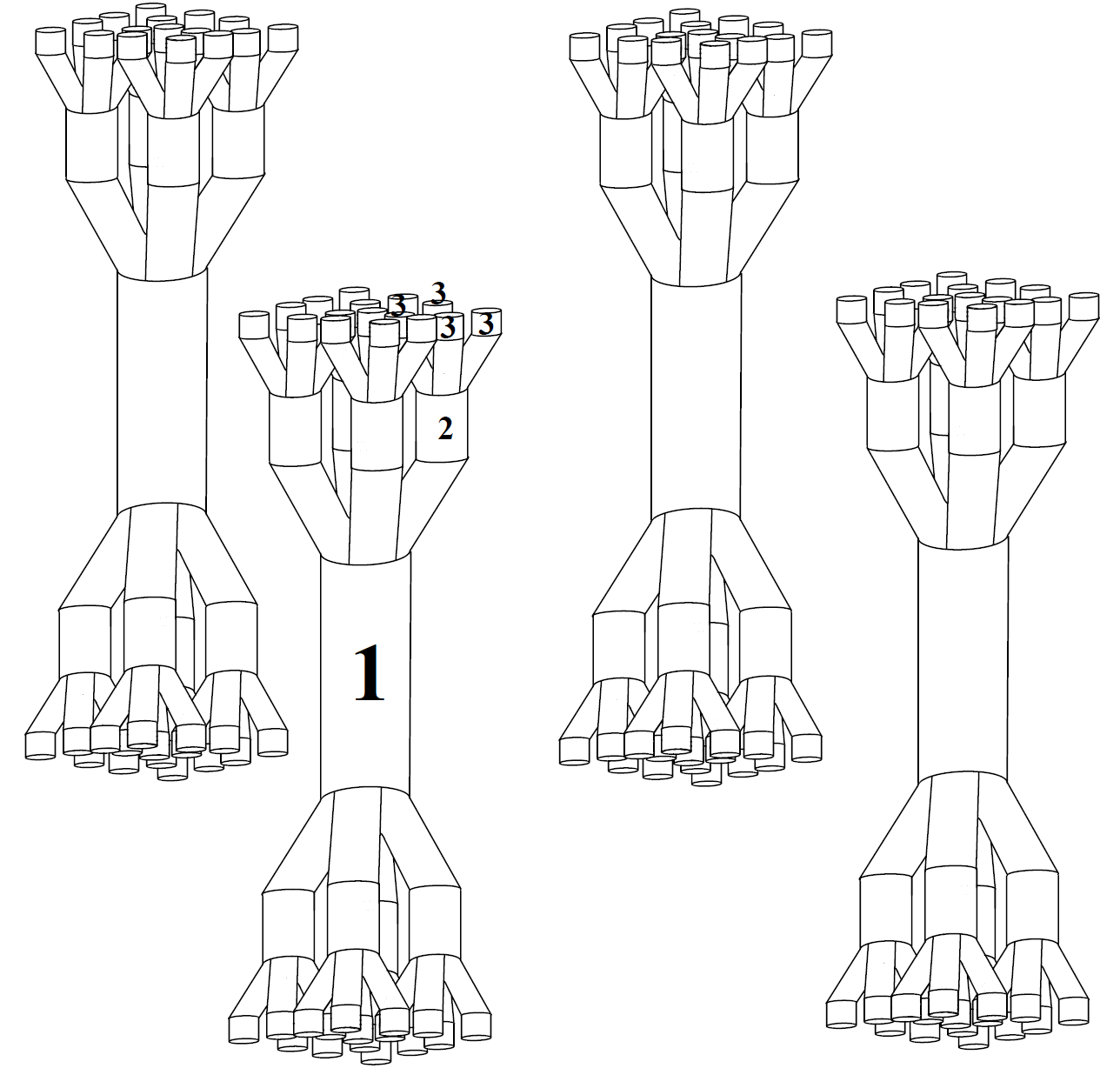}
    	\caption{}
    	\label{fig:3a}
    \end{subfigure}
    % \hfill
    \vspace{3mm}
    \begin{subfigure}[b]{0.4\textwidth}
    	\includegraphics[width=\textwidth]{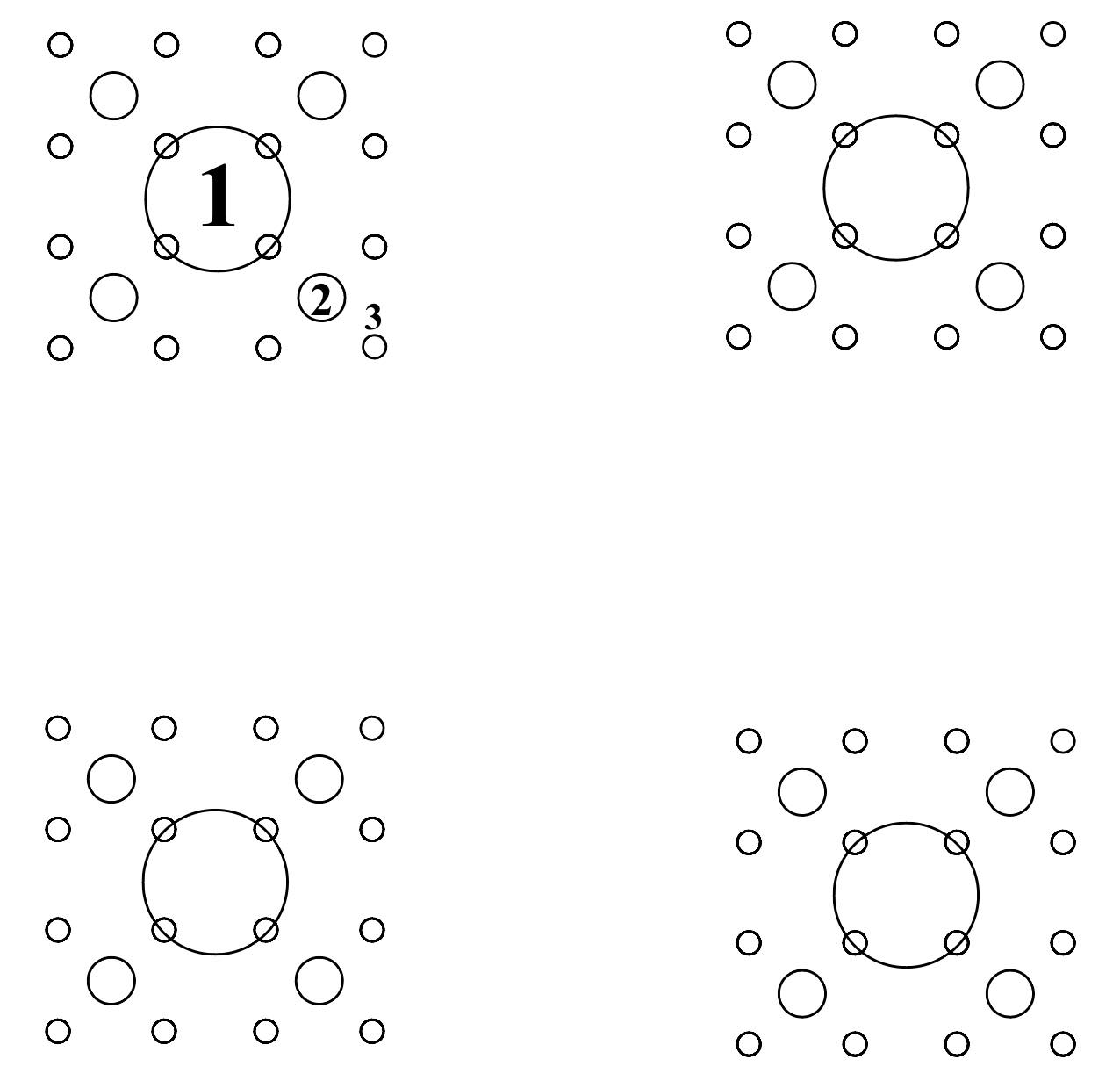}
    	\caption{}
    	\label{fig:3b}
    \end{subfigure}
    \hfill
    %\vspace{3mm}
    \caption{Schematic diagrams as in Figure \ref{fig:1}, but with four flux trees to illustrate the clustering of the smallest flux tubes (labelled by the numeral 3) within the same flux tree.}
    \label{fig:3}
\end{figure}

Figure \ref{fig:3}  illustrates what the foregoing results imply about the large-scale geometric arrangement of the flux tubes. 
The top and bottom panels of Figure \ref{fig:3} are the same as in Figure \ref{fig:1} but for four $N = 3$ flux trees.
It is apparent visually that quantized flux tubes cluster about the same flux tree with $d_\Phi < d_{\Phi, {\rm nb}}$, leaving regions in the type-II superconductor containing no quantized flux tubes, e.g.\ in the vacant gaps between the four groups of flux tubes in Figure \ref{fig:3b}. 
Since $\lambda$ decreases monotonically with increasing density \citep{glampedakis_magnetohydrodynamics_2011,link_instability_2012}, equation (\ref{eq:F_int2}) tells us that $F_{\rm  int}$ also decreases monotonically with increasing density for $\lambda < 2d_\Phi/3$.
That is, flux-tube repulsion is stronger at the type-II region far away from the type-I-II interface compared to the type-I-II interface, as one would expect. We do not assume a particular equation of state in this paper and make the conservative assumption that flux-tube repulsion is constant throughout the type-II region, where $\lambda$ in $F_{\rm t}$ corresponds to the value at $\rho \approx 2.8 \times 10^{14}$ g cm$^{-3}$. In other words, we overestimate flux-tube repulsion. Despite this, we find that quantized flux tubes cluster about the same flux tree.

\subsection{Flux tube clustering caused by superfluid-superconductor coupling: an alternative mechanism}
\label{sec:3.2}
Flux tube clustering in neutron stars has been predicted elsewhere to arise from a separate mechanism unrelated to the flux branching mechanism in Section \ref{sec:3.1} \citep{haber_critical_2017,wood_superconducting_2022}.
Specifically, it is argued that flux tube clustering arises in neutron stars due to the strong coupling between the neutron superfluid and proton superconductor, a phenomenon related to ``type-1.5'' superconductivity \citep{haber_critical_2017,wood_superconducting_2022}. 
Type-1.5 superconductivity was discovered experimentally in two-component terrestrial superconductors \citep{moshchalkov_type-15_2009,silaev_microscopic_2011}.
A type-1.5 superconductor has two coherence lengths, $\xi_1$ and $\xi_2$, and one penetration depth, $\lambda$, with $\xi_1 < \lambda < \xi_2$. 
The ordering of the length scales means that type-1.5 superconductors resemble both type-I and -II superconductors. 
Quantized flux tubes in type-1.5 superconductivity (with core size $\sim \xi_1$) attract mutually when their separation distance, $d_{\Phi, {\rm nb}} $, exceeds $\xi_2$. 
Long-range attraction also occurs between the mutually-attractive flux tubes in the intermediate states of type-I superconductivity \citep{kramer_thermodynamic_1971,alford_flux_2008}. 
On the other hand, for $d_{\Phi, {\rm nb}}  \sim \lambda$, quantized flux tubes in type-1.5 superconductivity repel mutually, just like the quantized flux tubes in type-II superconductivity. 
The competition between the repulsive $\lambda$ and attractive $\xi_2$ leads to a preferred intermediate length scale between quantized flux tubes, forming clusters of quantized flux tubes in type-1.5 superconductivity.
Quantized flux tubes in type-1.5 superconductivity approach each other but never coalesce in the flux tube region\footnote{\label{footnote:3}In a small part of the neutron star's parameter space, an intermediate state exists in type-1.5 superconductivity, such that flux tube lattice regions are interspersed with normal conducting regions, and flux tubes effectively merge \citep{wood_superconducting_2022}.}. 

We do not model the neutron superfluid in Section \ref{sec:3.1}, as the flux branching calculation is complicated enough without the neutrons.
Consequently, $d_\Phi$ obtained in Section \ref{sec:3.1} is likely to be greater than the realistic $d_\Phi$, when the neutron superfluid is included, and type-1.5 behavior becomes possible.

\section{Gravitational radiation}
\label{sec:4}
In this section, we study the astrophysical implications of flux tube clustering on the persistent gravitational radiation emitted by a neutron star.
The neutron superfluid inside the star rotates with $N_{\rm v} \sim 10^{17} (f/30{\rm Hz})$ quantized vortices, such that the superfluid mimics macroscopically rigid-body rotation, where each vortex carries a quantum of circulation $\kappa = \pi \hbar/m_n = 1.98 \times 10^{-3} \, {\rm cm^2 \, s^{-1}}$, $f$ is the neutron star's rotational frequency and $m_n$ is the mass of a neutron.
The vortices repel mutually and form an Abrikosov lattice in the absence of pinning.
In the outer core, vortices pin strongly to quantized flux tubes \citep{muslimov_vortex_1985,sauls_superfluidity_1989,srinivasan_novel_1990}, forming a non-axisymmetric vortex distribution which generates persistent, quasimonochromatic, current quadrupole gravitational radiation \citep{melatos_persistent_2015,haskell_continuous_2022,de_lillo_probing_2023,cheunchitra_persistent_2024}\footnote{The vortex distribution is not axisymmetric locally, even though it is approximately axisymmetric globally, when averaged over macroscopic length scales; see Figures 1 in both \citet{cheunchitra_persistent_2024} and \citet{melatos_persistent_2015}, for example.}. 
The characteristic gravitational wave strain, $h_0$, depends on the distribution of the quantized flux tubes, which serve as pinning sites.
\citet{cheunchitra_persistent_2024} considered pinning sites distributed homogeneously over a length-scale longer than the typical vortex separation, $d_{\rm v} \sim 2 \times 10^{-3} (f/30)^{1/2} \, {\rm cm}$. 
In contrast, the results in Section \ref{sec:3.1} suggest that the homogeneity assumption is not always valid, as vortices tend to cluster around flux trees, which may have $w > d_{\rm v}$.

In Section \ref{sec:4.1}, we review briefly the formalism to calculate $h_0$ for continuous current quadrupole gravitational radiation from an array of rectilinear superfluid vortices.
In Section \ref{sec:4.2}, we model the inhomogeneous vortex clustering about flux trees using a Matérn cluster point process to approximate $h_0$.
We emphasize that the calculations in this section are not fully self-consistent. The neutron superfluid is a passive participant. Physical effects arising from neutron-proton coupling, such as the type-1.5 superconductivity discussed in Section \ref{sec:3.2}, are therefore excluded from the calculation \citep{haber_critical_2017,wood_superconducting_2022}.
Additionally, superfluid vortices cluster around flux trees without affecting the structure of the flux trees. In other words, the neutron superfluid does not modify the state of the proton superconductor self-consistently, unlike in other work based on the coupled, time-dependent, Gross-Pitaevskii and Ginzburg-Landau equations \citep{thong_stability_2023,shukla_neutron-superfluid_2024}.
Relaxing these simplifications lies outside the scope of this paper and is postponed to future work. 
We also follow \citet{cheunchitra_persistent_2024} and postpone the study of tangled (as opposed to rectilinear) neutron superfluid vortices \citep{peralta_transitions_2006,andersson_superfluid_2007,drummond_stability_2017,drummond_stability_2018}.

\subsection{Current quadrupole moment}
\label{sec:4.1}
In the transverse traceless gauge, the far-field metric perturbation generated by the time-varying current quadrupole moment of a rectilinear array of quantized vortices in a neutron star with angular velocity $\Omega = 2\pi f$ and radius $R_*$ is given by $h^{\rm TT}_{jk}= T_{jk}^{{\rm B2},2,\pm1} h_0 \cos(\Omega t + \varphi)$ \citep{thorne_multipole_1980,warszawski_gravitational-wave_2012,melatos_persistent_2015,haskell_continuous_2022,de_lillo_probing_2023,cheunchitra_persistent_2024}, where $t$ is the retarded time, $\varphi$ is a phase that depends on the vortex position in the frame corotating with the star, and $T_{jk}^{{\rm B2}, 2\pm1}$ is the angular beam pattern emitted by the $(2,\pm 1)$ multipole moment, which depends on the observer's orientation relative to the source, defined in equation (2.30f) in \citet{thorne_multipole_1980}.
Following \citet{cheunchitra_persistent_2024}, we have
\begin{equation}
\label{eq:h_0}
    h_0 = \left(\frac{512\pi}{405}\right)^{1/2}\frac{G\Omega^2\rho \kappa R_*^4}{Dc^5} Q,
\end{equation}
where $\rho$ is the density of the neutron superfluid, which is assumed to be incompressible, $D$ is the unperturbed spatial distance between the star and Earth, $G$ is the gravitational constant, $c$ is the speed of light, and one has
\begin{align}
\label{eq:Q}
    Q = & \ \Biggl\{\Biggl[\sum_{i=1}^{N_{\rm v}} \frac{R_i}{R_*}\Biggl(1-\frac{R_i^2}{R_*^2} \Biggr)^{3/2}\cos\phi_{i,0}\Biggr]^2 \nonumber \\
    &+\Biggl[\sum_{i=1}^{N_{\rm v}} \frac{R_i}{R_*}\left(1-\frac{R_i^2}{R_*^2} \right)^{3/2}\sin\phi_{i,0}\Biggr]^2\Biggr\}^{1/2},
\end{align}
where the sum is over the equatorial positions $(R_i, \phi_{i, 0})$ in polar coordinates of the vortices in the corotating frame of the star.

\citet{cheunchitra_persistent_2024} estimated the median of $Q$ for a uniform Poisson distribution of vortices as
\begin{equation}
\label{eq:Q_old}
    Q_{\rm med} = 0.18 {N_{\rm v}}^{1/2},
\end{equation}
implying
\begin{align}
\label{eq:h_0_old}
    h_0 = & \ 1.2 \times 10^{-32} \left(\frac{f}{30 \, {\rm Hz}}\right)^{5/2} \nonumber \\
    &\times \left(\frac{R_*}{10 \, {\rm km}}\right)^{2} \left(\frac{M}{1.4 \, M_\odot}\right) \left(\frac{D}{1 \, {\rm kpc}}\right)^{-1}.
\end{align}
Equation (\ref{eq:h_0_old}) is an upper bound, when the mutual repulsion between vortices is negligible compared to the pinning energy, which leads to a Poissonian distribution of vortices. Numerical solutions demonstrate that the expected $h_0$ is likely to be orders of magnitude lower, with 
\begin{align}
\label{eq:h_0_old_num}
    h_0 = & \ 7.3 \times 10^{-42} \left(\frac{f}{30 \, {\rm Hz}}\right)^{5/2} \nonumber \\
    &\times \left(\frac{R_*}{10 \, {\rm km}}\right)^{2} \left(\frac{M}{1.4 \, M_\odot}\right) \left(\frac{D}{1 \, {\rm kpc}}\right)^{-1},
\end{align}
because mutual repulsion pushes vortices into a more regular array, which becomes approximately periodic, when the mutual repulsion dominates. As the array becomes more periodic, additional cancellation occurs between positive and negative contributions to the integrand of the current quadrupole moment, which is the source of the gravitational radiation studied here;
see Section 6 in \citet{cheunchitra_persistent_2024} for a thorough discussion.

\subsection{Vortex clustering}
\label{sec:4.2}
Vortex clustering, e.g.\ the arrangement illustrated in Figure \ref{fig:3}, modifies the spatially homogeneous Poisson point process assumed when deriving (11) \citep{cheunchitra_persistent_2024}. 
It is challenging to calculate the moments of $Q$ and hence estimate $h_0$ exactly, when vortex clustering occurs, because the geometry in Figure \ref{fig:3} does not possess a straightforward analytic form. 
Instead, by way of approximation, we assume that all vortices are rectilinear and cluster around flux trees with the same branching level $N$ and radius $r_{\rm ft} \sim w$.
We model the position of the centers of flux trees (e.g.\ the centres of the four groups of circles in Figure \ref{fig:3b}) as a uniform Poisson distribution with average flux tree count $N_{\rm ft} = N_{\rm f}/4^{N-1}$, where $N_{\rm f} \sim 10^{30} (B/10^{12} \, {\rm G})$ is the total number of flux tubes in a typical neutron star. 
In reality, due to macroscopic flux freezing \citep{baym_superfluidity_1969}, the spatial distribution of flux trees is complicated and linked intricately to the star's formation history, the study of which is outside the scope of this paper (see footnote \ref{footnote:3}).
Within each flux tree, e.g.\ within each of the four groups of circles in Figure \ref{fig:3b}, we assume that pinned vortices are distributed according to a different uniform Poisson process, with average vortex count per flux tree $N_{\rm v,t} = N_{\rm v} / N_{\rm ft}$.
That is, the vortex positions are modelled using a Matérn cluster point process \citep{baddeley_spatial_2015}. A Mat\'{e}rn cluster point process process is composed of two nested Poisson processes. Specifically, $N_{\rm ft}$ parent points (the centers of flux trees) are uniformly Poisson distributed in the equatorial plane, and $N_{\rm v, t}$ offspring points (vortices) are uniformly Poisson distributed in a circle with radius $r_{\rm ft}$ centered on each parent.

We use Campbell's theorem and a modified central limit theorem \citep{kulperger_central_1987,baddeley_spatial_2015} to analytically derive the median of $Q$, denoted by $Q_{\rm med}$, in Appendix \ref{appendix:A} for the Mat\'{e}rn cluster point process above. The result is 
\begin{equation}
\label{eq:Q_new}
    Q_{\rm med} = 0.18 N_{\rm v}^{1/2} (1 + N_{\rm v, t})^{1/2},
\end{equation}
which implies
\begin{align}
\label{eq:h_0_new1}
    h_0 = & \ 1.2 \times 10^{-32}  \left(1+N_{\rm v, t}\right)^{1/2}  \left(\frac{f}{30 \, {\rm Hz}}\right)^{5/2} \nonumber \\
    &\times \left(\frac{R_*}{10 \, {\rm km}}\right)^{2} \left(\frac{M}{1.4 \, M_\odot}\right) \left(\frac{D}{1 \, {\rm kpc}}\right)^{-1},
\end{align}
with $1 \leq N_{\rm v, t} \lesssim N_{\rm v} \sim  10^{17} (f/30{\rm Hz})$.
Thus, clustered vortices are expected to produce a median wave strain $h_0$, which is $(1+N_{\rm v, t})^{1/2}$ times larger than the $h_0$ produced by a uniform Poisson distribution of vortices.

If the star contains relatively few (and hence relatively large) flux trees, the enhancement in $h_0$ can reach astronomically significant levels. As an illustrative example, let us take $N_{\rm ft} \sim 10^3$. This estimate follows from magnetic flux conservation, if one substitutes $L_1 \sim 10^6 \, {\rm cm}$, $a_1 \sim 10^3 \, {\rm cm}$, and $B/H_{\rm c} \sim 10^{-3}$ into equation (18) for $N_{\rm ft}$ in \citet{thong_magnetic_2024}, viz.\
\begin{equation}
\label{eq:Nft}
    N_{\rm ft} = \frac{B (L_1/2)^2}{H_{\rm c} a_1^2}.
\end{equation}
Then one obtains $h_0 \sim 10^{-25}$ for the fiducial values of $f$, $R_\ast$, $M$, and $D$ quoted in (\ref{eq:h_0_new1}). Wave strains of this magnitude are marginally detectable as quasimonochromatic continuous wave signals by the current generation of long-baseline interferometers like the Laser Interferometer Gravitational Wave Observatory (LIGO) \citep{riles_searches_2023,wette_searches_2023}.

\section{Conclusions}
In this paper, we predict that quantized flux tubes form clusters within the type-II proton superconductor in the outer core of a neutron star, when it is adjacent to and magnetically coupled with a type-I proton superconductor in the inner core.
Previous studies \citep{thong_magnetic_2024} find that macroscopic flux tubes in the type-I superconductor divide dendritically into quantized flux tubes in the type-II superconductor, a phenomenon referred to as flux branching.
We extend the free energy calculations of flux branching in \citet{thong_magnetic_2024}  by incorporating the magnetic repulsion between quantized flux tubes in the type-II superconductor.
By balancing the mutual repulsion against the flux tree's tendency to minimize its width, $w$, we determine the minimum-energy separation, $d_\Phi$, between quantized flux tubes in the same flux tree to be two to seven times less than the separation, $d_{\Phi, {\rm nb}} \sim 5 \times 10^{-10} \, {\rm cm}$, without flux branching \citep{glampedakis_magnetohydrodynamics_2011}.
Specifically, we calculate the minimum-energy $(d_\Phi, w)$ values to be $(2.2 \times 10^{-10}\, \rm{cm}, 2.9 \times 10^{-2} \, \rm{cm}), (1.4\times 10^{-10} \, \rm{cm}, 1.9 \times 10^{2} \, \rm{cm})$ and $(6.5\times 10^{-11} \, \rm{cm}, 9.1 \times 10^{3} \, \rm{cm})$ for $N=28,38, $ and $48$, respectively. 
That is, quantized flux tubes cluster within their own flux tree, and the type-II superconductor is divided into regions with and without quantized flux tubes, as illustrated in Figure \ref{fig:3}.
For $N=38$, we calculate the minimum-energy $d_\Phi$ for $(L_1, L_2)/ (10^{6} \, {\rm cm}) = (0.01, 1.99), (1.99, 0.01)$ to be $(1.2, 1.2) \times 10^{-10} \, \rm{cm}$, respectively; flux clustering is insensitive to the radial thickness of the type-I and -II regions.
We note that flux tube clustering has also been proposed to arise from density and current-current coupling between the proton superconductor and the neutron superfluid, an alternative mechanism independent of flux branching \citep{haber_critical_2017,wood_superconducting_2022}.

In Section \ref{sec:4}, we show that in a highly idealized, non-dynamical model subject to the caveats itemized in Sections \ref{sec:2} and \ref{sec:3}, flux tube clustering enhances substantially the upper bound of the persistent, quasimonochromatic gravitational radiation emitted by the array of neutron superfluid vortices inside a neutron star. 
Neutron vortices pin to quantized flux tubes non-axisymmetrically, emitting current quadrupole gravitational radiation \citep{melatos_persistent_2015,haskell_continuous_2022,de_lillo_probing_2023,cheunchitra_persistent_2024}. 
Without flux tube clustering, vortices form a rectilinear array distributed according to a uniform Poisson point process, a good approximation when pinning is stronger than vortex-vortex repulsion, if one neglects tangling.
With flux tube clustering, vortices cluster about flux trees and their positions can be modelled with a Mat\'{e}rn cluster point process.
In the latter scenario, the median $h_0$ is given by 
\begin{align}
\label{eq:final_result}
    h_0 = & \ 1.2 \times 10^{-32}  \left(1+N_{\rm v, t}\right)^{1/2}  \left(\frac{f}{30 \, {\rm Hz}}\right)^{5/2} \nonumber \\
    &\times \left(\frac{R_*}{10 \, {\rm km}}\right)^{2} \left(\frac{M}{1.4 \, M_\odot}\right) \left(\frac{D}{1 \, {\rm kpc}}\right)^{-1},
\end{align}
where $1 \leq N_{\rm v, t} \lesssim  10^{17} (f/30{\rm Hz})$ is the mean number of vortices pinned to each flux tree. 
For example, relatively large flux trees with branching level $N = 48$ and hence $N_{\rm v, t} \sim 2 \times 10^{15}$, lead to $h_0 \sim 5.3\times 10^{-25}$ for the fiducial values quoted above.
This result is important for gravitational wave astronomy, as it approaches the sensitivity limit applying to continuous wave signals with the current generation of interferometric gravitational wave detectors, such as LIGO. 

We emphasize that more work is needed to relax the idealizations identified throughout the paper, before definitive statements about the gravitational wave signal can be made. 
For example, in the paper we assume a Poissonian spatial distribution of flux trees. In reality, magnetic evolution of the star may regularize the flux tree distribution, e.g.\ through turbulent mixing shortly after birth \citep{tout_magnetic_2004,ferrario_magnetic_2015}, which likely leads to a smaller $h_0$. We also assume a Poissonian spatial distribution of vortices about each flux tree. 
If vortices are arranged in a lattice about each flux tree, one may expect a lower $h_0$ as implied by the arguments of \citet{cheunchitra_persistent_2024}. That said, the distribution of the parent points (flux trees) is arguably more important than that of the offspring points (vortices) in this context for $h_0$. Suppose that $N_{\rm v, t}$ vortices are in a lattice about each flux tree. 
One may approximate the contribution to the current quadropole of these $N_{\rm v, t}$ vortices as a giant vortex with vorticity $N_{\rm v, t}\kappa$ at the centre of the flux tree (assuming $r_{\rm ft} \ll R_*$). 
Then, by transforming $\kappa \mapsto N_{\rm v, t} \kappa$ in equation (\ref{eq:h_0}) and $N_{\rm v} \mapsto N_{\rm ft}$ in equation (\ref{eq:Q_old}), the modified strain is given by equation (\ref{eq:h_0_old}) times a factor of $N_{\rm v, t}^{1/2}$.
In other words, if vortices form a lattice in each flux tree, then one can approximate the current quadropole of this configuration as $N_{\rm ft}$ Poisson-distributed large vortices, each with vorticity $N_{\rm v,t} \kappa$, consistent with (\ref{eq:final_result}). Additionally, results from \citet{haber_critical_2017} and \citet{wood_superconducting_2022} indicate that flux tubes are likely to be more clustered than just from flux branching alone, implying more flux tubes per cluster. For a given total flux, this reduces the number of clusters for vortices to pin to and hence increases $N_{\rm v, t}$, leading to a higher $h_0$ in equation (\ref{eq:final_result}). Quantifying the net outcome of these competing effects lies outside the scope of this paper.

This paper makes several other simplifying assumptions. 
(i) The sharp-interface approximation \citep{landau_theory_1943,andrew_intermediate_1948,huebener_magnetic_2001} should be replaced more generally with the time-dependent Ginzburg-Landau equation. Indeed, more generally, the Ginzburg-Landau equation should be solved simultaneously with the Gross-Pitaevskii equation, to treat self-consistently the coupling between the proton superconductor and neutron superfluid \citep{thong_stability_2023,shukla_neutron-superfluid_2024} and accommodate important effects such as type-1.5 superconductivity \citep{haber_critical_2017,wood_superconducting_2022}. 
(ii) The radius, $a_1$, of the largest macroscopic flux tubes (trunks) is the same for all flux trees, whereas in reality, $a_1$ varies from one flux tree to the next depending on the magnetic history of the star, especially any dynamo activity early in its life \citep{tout_magnetic_2004,ferrario_magnetic_2015}.
(iii) All vortices in the outer core are pinned to flux trees, overlooking that vortices can pin to isolated quantized flux tubes in the type-II superconductor that are not part of any flux trees \citep{thong_magnetic_2024}\footnote{Due to the spherical geometry of the star, quantized flux tubes in the type-II region can avoid the type-I region entirely, i.e., lines threading the outer core may not cross the inner core. Such flux tubes are not part of any flux trees and do not cluster \citep{thong_magnetic_2024}.}.
(iv) Vortex clustering about flux trees is modelled by a Mat\'{e}rn cluster point process, a statistical assumption which is reasonable but cannot be tested experimentally under neutron star conditions now or in the foreseeable future.
(v) The inner core is entirely a type-I proton superconductor, whereas in reality, it may coexist with other phases of dense matter, such as hypersonic matter \citep{vidana_hyperon-hyperon_2000,nishizaki_hyperon-mixed_2002,schaffner-bielich_phase_2002} and color superconductivity \citep{alford_magnetic_2000,alford_color-superconducting_2001,lattimer_physics_2004,alford_color-magnetic_2010,baym_hadrons_2018}. 
In the latter scenario, however, the paper's conclusion still holds qualitatively and partially, if regions of type-I and -II proton superconductivity exist somewhere adjacent to one another.

\section*{Acknowledgements}

The authors thank the anonymous referee for insightful comments, including for drawing our attention to the property in footnote \ref{footnote:3}.The authors thank Yongyan Xia for helping to draw Figures \ref{fig:1} and \ref{fig:2} and acknowledge discussions with Thippayawis Cheunchitra and Julian Carlin. Parts of this research are supported by an Australian Government Research Training Program Scholarship (Stipend), Research Training Program Scholarship (Fee Offset), Rowden White Scholarship, McKellar Prize in Theoretical Physics, Professor Kernot Research Scholarship in Physics and the Australian Research Council (ARC) Centre of Excellence for Gravitational Wave Discovery (OzGrav) (grant number CE170100004).

%%%%%%%%%%%%%%%%%%%%%%%%%%%%%%%%%%%%%%%%%%%%%%%%%%
\section*{Data Availability}

 No new data were generated or analysed in support of this research.



\bibliographystyle{mnras}
\bibliography{references} % if your bibtex file is called example.bib

\begin{thebibliography}{}
\makeatletter
\relax
\def\mn@urlcharsother{\let\do\@makeother \do\$\do\&\do\#\do\^\do\_\do\%\do\~}
\def\mn@doi{\begingroup\mn@urlcharsother \@ifnextchar [ {\mn@doi@} {\mn@doi@[]}}
\def\mn@doi@[#1]#2{\def\@tempa{#1}\ifx\@tempa\@empty \href {http://dx.doi.org/#2} {doi:#2}\else \href {http://dx.doi.org/#2} {#1}\fi \endgroup}
\def\mn@eprint#1#2{\mn@eprint@#1:#2::\@nil}
\def\mn@eprint@arXiv#1{\href {http://arxiv.org/abs/#1} {{\tt arXiv:#1}}}
\def\mn@eprint@dblp#1{\href {http://dblp.uni-trier.de/rec/bibtex/#1.xml} {dblp:#1}}
\def\mn@eprint@#1:#2:#3:#4\@nil{\def\@tempa {#1}\def\@tempb {#2}\def\@tempc {#3}\ifx \@tempc \@empty \let \@tempc \@tempb \let \@tempb \@tempa \fi \ifx \@tempb \@empty \def\@tempb {arXiv}\fi \@ifundefined {mn@eprint@\@tempb}{\@tempb:\@tempc}{\expandafter \expandafter \csname mn@eprint@\@tempb\endcsname \expandafter{\@tempc}}}

\bibitem[\protect\citeauthoryear{Alford}{Alford}{2001}]{alford_color-superconducting_2001}
Alford M.,  2001, \mn@doi [Annual Review of Nuclear and Particle Science] {10.1146/annurev.nucl.51.101701.132449}, 51, 131

\bibitem[\protect\citeauthoryear{Alford \& Good}{Alford \& Good}{2008}]{alford_flux_2008}
Alford M.~G.,  Good G.,  2008, \mn@doi [Physical Review B] {10.1103/PhysRevB.78.024510}, 78, 024510

\bibitem[\protect\citeauthoryear{Alford \& Sedrakian}{Alford \& Sedrakian}{2010}]{alford_color-magnetic_2010}
Alford M.~G.,  Sedrakian A.,  2010, \mn@doi [Journal of Physics G: Nuclear and Particle Physics] {10.1088/0954-3899/37/7/075202}, 37, 075202

\bibitem[\protect\citeauthoryear{Alford, Berges  \& Rajagopal}{Alford et~al.}{2000}]{alford_magnetic_2000}
Alford M.,  Berges J.,   Rajagopal K.,  2000, \mn@doi [Nuclear Physics B] {10.1016/S0550-3213(99)00830-5}, 571, 269

\bibitem[\protect\citeauthoryear{Andersson, Sidery  \& Comer}{Andersson et~al.}{2007}]{andersson_superfluid_2007}
Andersson N.,  Sidery T.,   Comer G.~L.,  2007, \mn@doi [Monthly Notices of the Royal Astronomical Society] {10.1111/j.1365-2966.2007.12251.x}, 381, 747

\bibitem[\protect\citeauthoryear{Andrew}{Andrew}{1948}]{andrew_intermediate_1948}
Andrew E.~R.,  1948, \mn@doi [Proceedings of the Royal Society of London Series A] {10.1098/rspa.1948.0068}, 194, 98

\bibitem[\protect\citeauthoryear{Baddeley, Rubak  \& Turner}{Baddeley et~al.}{2015}]{baddeley_spatial_2015}
Baddeley A.,  Rubak E.,   Turner R.,  2015, Spatial {Point} {Patterns}: {Methodology} and {Applications} with {R}.
CRC Press

\bibitem[\protect\citeauthoryear{Baym, Pethick  \& Pines}{Baym et~al.}{1969a}]{baym_superfluidity_1969}
Baym G.,  Pethick C.,   Pines D.,  1969a, \mn@doi [Nature] {10.1038/224673a0}, 224, 673

\bibitem[\protect\citeauthoryear{Baym, Pethick  \& Pikes}{Baym et~al.}{1969b}]{baym_electrical_1969}
Baym G.,  Pethick C.,   Pikes D.,  1969b, \mn@doi [Nature] {10.1038/224674a0}, 224, 674

\bibitem[\protect\citeauthoryear{Baym, Hatsuda, Kojo, Powell, Song  \& Takatsuka}{Baym et~al.}{2018}]{baym_hadrons_2018}
Baym G.,  Hatsuda T.,  Kojo T.,  Powell P.~D.,  Song Y.,   Takatsuka T.,  2018, \mn@doi [Reports on Progress in Physics] {10.1088/1361-6633/aaae14}, 81, 056902

\bibitem[\protect\citeauthoryear{Buckley, Metlitski  \& Zhitnitsky}{Buckley et~al.}{2004}]{buckley_neutron_2004}
Buckley K. B.~W.,  Metlitski M.~A.,   Zhitnitsky A.~R.,  2004, \mn@doi [Physical Review Letters] {10.1103/PhysRevLett.92.151102}, 92, 151102

\bibitem[\protect\citeauthoryear{Chapman}{Chapman}{1995}]{chapman_asymptotic_1995}
Chapman S.~J.,  1995, \mn@doi [Quarterly of Applied Mathematics] {10.1090/qam/1359498}, 53, 601

\bibitem[\protect\citeauthoryear{Cheunchitra, Melatos, Carlin  \& Howitt}{Cheunchitra et~al.}{2024}]{cheunchitra_persistent_2024}
Cheunchitra T.,  Melatos A.,  Carlin J.~B.,   Howitt G.,  2024, \mn@doi [Monthly Notices of the Royal Astronomical Society] {10.1093/mnras/stae130}, 528, 1360

\bibitem[\protect\citeauthoryear{Choksi, Kohn  \& Otto}{Choksi et~al.}{2004}]{choksi_energy_2004}
Choksi R.,  Kohn R.~V.,   Otto F.,  2004, \mn@doi [Journal of Nonlinear Science] {10.1007/s00332-004-0568-2}, 14, 119

\bibitem[\protect\citeauthoryear{De~Lillo, Suresh, Depasse, Sieniawska, Miller  \& Bruno}{De~Lillo et~al.}{2023}]{de_lillo_probing_2023}
De~Lillo F.,  Suresh J.,  Depasse A.,  Sieniawska M.,  Miller A.~L.,   Bruno G.,  2023, \mn@doi [Physical Review D] {10.1103/PhysRevD.107.102001}, 107, 102001

\bibitem[\protect\citeauthoryear{Drummond \& Melatos}{Drummond \& Melatos}{2017}]{drummond_stability_2017}
Drummond L.~V.,  Melatos A.,  2017, \mn@doi [Monthly Notices of the Royal Astronomical Society] {10.1093/mnras/stx2301}, 472, 4851

\bibitem[\protect\citeauthoryear{Drummond \& Melatos}{Drummond \& Melatos}{2018}]{drummond_stability_2018}
Drummond L.~V.,  Melatos A.,  2018, \mn@doi [Monthly Notices of the Royal Astronomical Society] {10.1093/mnras/stx3197}, 475, 910

\bibitem[\protect\citeauthoryear{Ferrario, Melatos  \& Zrake}{Ferrario et~al.}{2015}]{ferrario_magnetic_2015}
Ferrario L.,  Melatos A.,   Zrake J.,  2015, \mn@doi [Space Science Reviews] {10.1007/s11214-015-0138-y}, 191, 77

\bibitem[\protect\citeauthoryear{Glampedakis, Andersson  \& Samuelsson}{Glampedakis et~al.}{2011}]{glampedakis_magnetohydrodynamics_2011}
Glampedakis K.,  Andersson N.,   Samuelsson L.,  2011, \mn@doi [Monthly Notices of the Royal Astronomical Society] {10.1111/j.1365-2966.2010.17484.x}, 410, 805

\bibitem[\protect\citeauthoryear{Haber \& Schmitt}{Haber \& Schmitt}{2017}]{haber_critical_2017}
Haber A.,  Schmitt A.,  2017, \mn@doi [Physical Review D] {10.1103/PhysRevD.95.116016}, 95, 116016

\bibitem[\protect\citeauthoryear{Haskell, Pizzochero  \& Seveso}{Haskell et~al.}{2013}]{haskell_investigating_2013}
Haskell B.,  Pizzochero P.~M.,   Seveso S.,  2013, \mn@doi [The Astrophysical Journal] {10.1088/2041-8205/764/2/L25}, 764, L25

\bibitem[\protect\citeauthoryear{Haskell, Antonelli  \& Pizzochero}{Haskell et~al.}{2022}]{haskell_continuous_2022}
Haskell B.,  Antonelli M.,   Pizzochero P.,  2022, \mn@doi [Universe] {10.3390/universe8120619}, 8, 619

\bibitem[\protect\citeauthoryear{Hubert}{Hubert}{1967}]{hubert_theory_1967}
Hubert A.,  1967, \mn@doi [Phys. Status Solidi] {10.1002/pssb.19670240229}, 24

\bibitem[\protect\citeauthoryear{Huebener}{Huebener}{2001}]{huebener_magnetic_2001}
Huebener R.~P.,  2001, Magnetic {Flux} {Structures} in {Superconductors}: {Extended} {Reprint} of a {Classic} {Text}.
 Springer {Series} in {SOLID}-{STATE} {SCIENCES} Vol. 6, Springer, Berlin, Heidelberg, \mn@doi{10.1007/978-3-662-08446-5}, \url {https://link.springer.com/10.1007/978-3-662-08446-5}

\bibitem[\protect\citeauthoryear{Jones}{Jones}{2006}]{jones_type_2006}
Jones P.~B.,  2006, \mn@doi [Monthly Notices of the Royal Astronomical Society] {10.1111/j.1365-2966.2006.10754.x}, 371, 1327

\bibitem[\protect\citeauthoryear{Kramer}{Kramer}{1971}]{kramer_thermodynamic_1971}
Kramer L.,  1971, \mn@doi [Physical Review B] {10.1103/PhysRevB.3.3821}, 3, 3821

\bibitem[\protect\citeauthoryear{Kulperger}{Kulperger}{1987}]{kulperger_central_1987}
Kulperger R.~J.,  1987, in MacNeill I.~B.,  Umphrey G.~J.,  Bellhouse D.~R.,   Kulperger R.~J.,  eds, , Advances in the {Statistical} {Sciences}: {Applied} {Probability}, {Stochastic} {Processes}, and {Sampling} {Theory}: {Volume} {I} of the {Festschrift} in {Honor} of {Professor} {V}.{M}. {Joshi}’s 70th {Birthday}.
Springer Netherlands, Dordrecht, pp 131--139, \mn@doi{10.1007/978-94-009-4786-3_10}, \url {https://doi.org/10.1007/978-94-009-4786-3_10}

\bibitem[\protect\citeauthoryear{Landau}{Landau}{1943}]{landau_theory_1943}
Landau L.,  1943, J. Phys. U.S.S.R., 7, 99

\bibitem[\protect\citeauthoryear{Lattimer \& Prakash}{Lattimer \& Prakash}{2004}]{lattimer_physics_2004}
Lattimer J.~M.,  Prakash M.,  2004, \mn@doi [Science] {10.1126/science.1090720}, 304, 536

\bibitem[\protect\citeauthoryear{Link}{Link}{2012}]{link_instability_2012}
Link B.,  2012, \mn@doi [Monthly Notices of the Royal Astronomical Society] {10.1111/j.1365-2966.2012.20498.x}, 421, 2682

\bibitem[\protect\citeauthoryear{Melatos, Douglass  \& Simula}{Melatos et~al.}{2015}]{melatos_persistent_2015}
Melatos A.,  Douglass J.~A.,   Simula T.~P.,  2015, \mn@doi [The Astrophysical Journal] {10.1088/0004-637X/807/2/132}, 807, 132

\bibitem[\protect\citeauthoryear{Mendell}{Mendell}{1991}]{mendell_superfluid_1991}
Mendell G.,  1991, \mn@doi [The Astrophysical Journal] {10.1086/170609}, 380, 515

\bibitem[\protect\citeauthoryear{Moshchalkov et~al.,}{Moshchalkov et~al.}{2009}]{moshchalkov_type-15_2009}
Moshchalkov V.,  et~al., 2009, \mn@doi [Physical Review Letters] {10.1103/PhysRevLett.102.117001}, 102, 117001

\bibitem[\protect\citeauthoryear{Muslimov \& Tsygan}{Muslimov \& Tsygan}{1985}]{muslimov_vortex_1985}
Muslimov A.~G.,  Tsygan A.~I.,  1985, \mn@doi [Astrophysics and Space Science] {10.1007/BF00653825}, 115, 43

\bibitem[\protect\citeauthoryear{Nishizaki, Yamamoto  \& Takatsuka}{Nishizaki et~al.}{2002}]{nishizaki_hyperon-mixed_2002}
Nishizaki S.,  Yamamoto Y.,   Takatsuka T.,  2002, \mn@doi [Progress of Theoretical Physics] {10.1143/PTP.108.703}, 108, 703

\bibitem[\protect\citeauthoryear{Peralta, Melatos, Giacobello  \& Ooi}{Peralta et~al.}{2006}]{peralta_transitions_2006}
Peralta C.,  Melatos A.,  Giacobello M.,   Ooi A.,  2006, \mn@doi [The Astrophysical Journal] {10.1086/507576}, 651, 1079

\bibitem[\protect\citeauthoryear{Riles}{Riles}{2023}]{riles_searches_2023}
Riles K.,  2023, \mn@doi [Living Reviews in Relativity] {10.1007/s41114-023-00044-3}, 26, 3

\bibitem[\protect\citeauthoryear{Sauls}{Sauls}{1989}]{sauls_superfluidity_1989}
Sauls J.~A.,  1989, in Ögelman H.,  van~den Heuvel E. P.~J.,  eds, {NATO} {ASI} {Series}, Timing {Neutron} {Stars}.
Springer Netherlands, Dordrecht, pp 457--490, \mn@doi{10.1007/978-94-009-2273-0_43}, \url {https://doi.org/10.1007/978-94-009-2273-0_43}

\bibitem[\protect\citeauthoryear{Schaffner-Bielich, Hanauske, Stocker  \& Greiner}{Schaffner-Bielich et~al.}{2002}]{schaffner-bielich_phase_2002}
Schaffner-Bielich J.,  Hanauske M.,  Stocker H.,   Greiner W.,  2002, \mn@doi [Physical Review Letters] {10.1103/PhysRevLett.89.171101}, 89, 171101

\bibitem[\protect\citeauthoryear{Sedrakian}{Sedrakian}{2005}]{sedrakian_type-i_2005}
Sedrakian A.,  2005, \mn@doi [Physical Review D] {10.1103/PhysRevD.71.083003}, 71, 083003

\bibitem[\protect\citeauthoryear{Sedrakian, Sedrakian  \& Zharkov}{Sedrakian et~al.}{1997}]{sedrakian_type_1997}
Sedrakian D.~M.,  Sedrakian A.~D.,   Zharkov G.~F.,  1997, \mn@doi [Monthly Notices of the Royal Astronomical Society] {10.1093/mnras/290.1.203}, 290, 203

\bibitem[\protect\citeauthoryear{Shukla, Brachet  \& Pandit}{Shukla et~al.}{2024}]{shukla_neutron-superfluid_2024}
Shukla S.,  Brachet M.~E.,   Pandit R.,  2024, Neutron-superfluid vortices and proton-superconductor flux tubes: {Development} of a minimal model for pulsar glitches, \mn@doi{10.48550/arXiv.2405.12127}, \url {http://arxiv.org/abs/2405.12127}

\bibitem[\protect\citeauthoryear{Silaev \& Babaev}{Silaev \& Babaev}{2011}]{silaev_microscopic_2011}
Silaev M.,  Babaev E.,  2011, \mn@doi [Physical Review B] {10.1103/PhysRevB.84.094515}, 84, 094515

\bibitem[\protect\citeauthoryear{Srinivasan, Bhattacharya, Muslimov  \& Tsygan}{Srinivasan et~al.}{1990}]{srinivasan_novel_1990}
Srinivasan G.,  Bhattacharya D.,  Muslimov A.~G.,   Tsygan A.~I.,  1990, Current Science, 59, 31

\bibitem[\protect\citeauthoryear{Thong \& Melatos}{Thong \& Melatos}{2024}]{thong_magnetic_2024}
Thong K.~H.,  Melatos A.,  2024, \mn@doi [Monthly Notices of the Royal Astronomical Society] {10.1093/mnras/stae2340}, 535, 551

\bibitem[\protect\citeauthoryear{Thong, Melatos  \& Drummond}{Thong et~al.}{2023}]{thong_stability_2023}
Thong K.~H.,  Melatos A.,   Drummond L.~V.,  2023, \mn@doi [Monthly Notices of the Royal Astronomical Society] {10.1093/mnras/stad927}, 521, 5724

\bibitem[\protect\citeauthoryear{Thorne}{Thorne}{1980}]{thorne_multipole_1980}
Thorne K.~S.,  1980, \mn@doi [Reviews of Modern Physics] {10.1103/RevModPhys.52.299}, 52, 299

\bibitem[\protect\citeauthoryear{Tinkham}{Tinkham}{2004}]{tinkham_introduction_2004}
Tinkham M.,  2004, Introduction to {Superconductivity}.
Courier Corporation

\bibitem[\protect\citeauthoryear{Tout, Wickramasinghe  \& Ferrario}{Tout et~al.}{2004}]{tout_magnetic_2004}
Tout C.~A.,  Wickramasinghe D.~T.,   Ferrario L.,  2004, \mn@doi [Monthly Notices of the Royal Astronomical Society] {10.1111/j.1365-2966.2004.08482.x}, 355, L13

\bibitem[\protect\citeauthoryear{Vidaña, Polls, Ramos, Engvik  \& Hjorth-Jensen}{Vidaña et~al.}{2000}]{vidana_hyperon-hyperon_2000}
Vidaña I.,  Polls A.,  Ramos A.,  Engvik L.,   Hjorth-Jensen M.,  2000, \mn@doi [Physical Review C] {10.1103/PhysRevC.62.035801}, 62, 035801

\bibitem[\protect\citeauthoryear{Warszawski \& Melatos}{Warszawski \& Melatos}{2012}]{warszawski_gravitational-wave_2012}
Warszawski L.,  Melatos A.,  2012, \mn@doi [Monthly Notices of the Royal Astronomical Society] {10.1111/j.1365-2966.2012.20977.x}, 423, 2058

\bibitem[\protect\citeauthoryear{Wette}{Wette}{2023}]{wette_searches_2023}
Wette K.,  2023, \mn@doi [Astroparticle Physics] {10.1016/j.astropartphys.2023.102880}, 153, 102880

\bibitem[\protect\citeauthoryear{Wood \& Graber}{Wood \& Graber}{2022}]{wood_superconducting_2022}
Wood T.~S.,  Graber V.,  2022, \mn@doi [Universe] {10.3390/universe8040228}, 8, 228

\makeatother
\end{thebibliography}

%\begin{thebibliography}{99}
%\end{thebibliography}

%%%%%%%%%%%%%%%%%%%%%%%%%%%%%%%%%%%%%%%%%%%%%%%%%%

%%%%%%%%%%%%%%%%% APPENDICES %%%%%%%%%%%%%%%%%%%%%

\appendix

\section{Analytic PDF of $Q$ for a Matérn cluster point process}
\label{appendix:A}
In this appendix, we calculate the probability distribution function (PDF) of $Q$ in (\ref{eq:Q}) for vortex positions $(R_i, \phi_{i, 0})$ situated randomly in the equatorial plane of the corotating frame according to a Matérn cluster point process, $\mathbf{X}$, on a two-dimensional region $S$, viz. a circular disk of radius $R_*$.

Suppose that the position of each flux tree axis is drawn independently from a uniform Poisson distribution with intensity $\Lambda_{\rm ft} = N_{\rm ft}/(\pi R_*^2)$. 
Each flux tree (or cluster) has a radius $r_{\rm ft} \sim w$, where $w$ is estimated by a minimum-energy argument in Section \ref{sec:3.1}.
Suppose that the vortex positions within a flux tree, i.e. within a disk of radius $r_{\rm ft}$ about the axis of a flux tree, are drawn independently from another uniform Poisson process with intensity $\Lambda_{\rm v, t} = N_{\rm v, t}/(\pi r_{\rm ft}^2)$.
The intensity, $\Lambda$, of the Matérn cluster point process, formed by combining the above two Poisson processes, is independent of position and is given by \citet{baddeley_spatial_2015}
\begin{align}
    \Lambda &= \Lambda_{\rm ft} N_{\rm v, t} \\
    &=\frac{N_{\rm ft} N_{\rm v, t}}{\pi R_*^2},
\end{align}
where $N_{\rm ft} N_{\rm v, t} = N_{\rm v}$ is the total number of vortices, i.e., $\Lambda$ is the same as that of a Poisson point process, whereas the second moment measure, which we get to below, is not the same.

We follow \citet{cheunchitra_persistent_2024} and re-express $Q$ from (\ref{eq:Q}) as
\begin{equation}
\label{eq:Q2}
    Q = \sqrt{S_x^2 + S_y^2},
\end{equation}
with
\begin{align}
    S_x &= \sum_{i=1}^{N_{\rm v}}W_{i,x}, \\ 
    S_y &= \sum_{i=1}^{N_{\rm v}}W_{i,y},
\end{align}
\begin{align}
    W_{i,x} &= \frac{R_i}{R_*}\Biggl(1-\frac{R_i^2}{R_*^2} \Biggr)^{3/2}\cos\phi_{i,0}, \\ 
    W_{i,y} &= \frac{R_i}{R_*}\Biggl(1-\frac{R_i^2}{R_*^2} \Biggr)^{3/2}\sin\phi_{i,0}.
\end{align}
To find the PDF of $Q$, we first find the PDFs of $S_x$ and $S_y$. 
\citet{cheunchitra_persistent_2024} treated $W_{i,x}$ and $W_{i,y}$ as independent and identically distributed variables, constructed from a Poisson distribution of vortex positions.
For a Matérn cluster point process (or a general Poisson cluster point process), vortex positions are not distributed independently.
In the regime $N_{\rm v} \gg 1$, we apply a modified central limit theorem \citep{kulperger_central_1987}, where the PDFs of $S_x$ and $S_y$ are normal distributions, whose means and variances differ from those of a Poisson process in general. 
Campbell's theorem implies \citep{baddeley_spatial_2015}
\begin{align}
    \langle S_x \rangle &= \int_0^{2 \pi} d\phi \,  \int_0^{R_*}  dR \, \Lambda R W_{x} \\
    &= 0,
\end{align}
with
\begin{equation}
    W_x = \frac{R}{R_*}\Biggl(1-\frac{R^2}{R_*^2} \Biggr)^{3/2}\cos\phi,
\end{equation}
and similarly $\langle S_y \rangle = 0$. Campbell's theorem also implies
\begin{align}
\label{eq:varS}
    \langle S_x^2 \rangle &= \int_S \int_{S'} \nu(dx, dx')\,    W_x W_{x'} , 
\end{align}
with $\nu(dx, dx') = \langle N_\mathbf{X}(dx) N_\mathbf{X}(dx') \rangle$, where $\nu$ is the second moment measure of $\mathbf{X}$, and $N_\mathbf{X}(dx) N_\mathbf{X}(dx')$ is the number of ordered pairs $(x, x')$ of points from the process $\mathbf{X}$ satisfying $x \in [x, x+dx]$ and $x' \in [x', x'+dx']$ \citep{baddeley_spatial_2015}. For a uniform Poisson point process, like that assumed by \citet{cheunchitra_persistent_2024}, we have $\nu(dx, dx') = \Lambda^2 dx dx' + \Lambda \delta(x-x')dxdx'$. For a Matérn cluster point process, we have
\begin{align}
\label{eq:nu}
    \nu(dx, dx') = & \  dx dx' \bigl\{\Lambda^2 + \Lambda \delta(x-x') \nonumber \\ &+ \Lambda_{\rm ft} \Lambda_{\rm v, t}^2 A\left[b(x, r_{\rm ft}) \cap b(x', r_{\rm ft})\right]\bigr\},
\end{align}
where $b(\alpha, \beta)$ is a circle centered on $\alpha$ with radius $\beta$ and $A$ is an area function.
That is, $A\left[b(x, r_{\rm ft}) \cap b(x', r_{\rm ft})\right]$ is the area of the intersection of two circles centered at $x$ and $x'$ with radii $r_{\rm ft}$.
Substituting (\ref{eq:nu}) into (\ref{eq:varS}), the first term is zero because we have $\langle S_x \rangle = 0$ while the second term simplifies to $N_{\rm ft} N_{\rm v, t}/40$; see Appendix B in \citet{cheunchitra_persistent_2024}.
The third term, $I_3$, is more complicated and has the form 
\begin{align}
\label{eq:I3}
    I_3 = \int_S \int_{S'} dx dx' \,    W_x W_{x'} \Lambda_{\rm ft} \Lambda_{\rm v, t}^2 A\left[b(x, r_{\rm ft}) \cap b(x', r_{\rm ft})\right]. 
\end{align}
In (\ref{eq:I3}), $A\left[b(x, r_{\rm ft}) \cap b(x', r_{\rm ft})\right]$ varies quickly with $x$ and $x'$ and vanishes for $|x - x'| > 2r_{\rm ft}$, while $W_x$ and $W_{x'}$ vary slowly given $r_{\rm ft} \ll R_*$. 
Hence, we can approximate $W_x \approx W_{x'}$ to obtain 
% In (\ref{eq:I3}), $A\left[b(x, r_{\rm ft}) \cap b(x', r_{\rm ft})\right]$ varies quickly with $x$ and $x'$, while $W_x$ and $W_{x'}$ vary slowly given $r_{\rm ft} \ll R_*$. 
% Hence, we approximate (\ref{eq:I3}) as
\begin{align}
\label{eq:I3_2}
    I_3 \approx \int_S dx \,    W_x W_{x} \Lambda_{\rm ft} \Lambda_{\rm v, t}^2 I_A, 
\end{align}
with 
\begin{align}
    I_A = \int_{S'} dx' \, A\left[b(x, r_{\rm ft}) \cap b(x', r_{\rm ft})\right],
\end{align}
where $I_A$ measures the overlap of the two circles.
For $S = S'$ equal to the equatorial cross-section of the star, we find $I_A = \pi^2 r_{\rm ft}^4$, and hence
\begin{equation}
    I_3 \approx \frac{N_{\rm ft} N_{\rm v, t}^2}{40} ,
\end{equation}
and
\begin{equation}
\label{eq:varS2}
    \langle S_x^2 \rangle \approx \frac{N_{\rm ft} N_{\rm v, t}}{40} (1 + N_{\rm v, t}).
\end{equation}
Similarly we have $\langle S_y^2 \rangle \approx N_{\rm ft} N_{\rm v, t}(1 + N_{\rm v, t})/40$.
That is, the PDFs of $S_x$ and $S_y$ are Gaussians with zero mean and variance given by (\ref{eq:varS2}).
From (\ref{eq:Q2}), the PDF of $Q$ simplfies to
\begin{equation}
    p(Q) = \frac{Q}{\langle S_x^2 \rangle
    } \exp\left(-\frac{Q^2}{2\langle S_x^2 \rangle}\right),
\end{equation}
with 
\begin{equation}
    Q_{\rm med} = 0.18 N_{\rm v}^{1/2} (1 + N_{\rm v, t})^{1/2}.
\end{equation}
%%%%%%%%%%%%%%%%%%%%%%%%%%%%%%%%%%%%%%%%%%%%%%%%%%

% Don't change these lines
\bsp	% typesetting comment
\label{lastpage}
\end{document}